\documentclass[preprint,showpacs,amsmath,amssymb]{revtex4}

\usepackage{graphicx}
\usepackage{dcolumn}
\usepackage{bm}

\begin{document}

\title{Anisotropic Hall Effect in Single Crystal Heavy Fermion YbAgGe }

\author{Sergey L. Bud'ko$^\ast$, Emilia Morosan$^{\ast\dagger}$, and Paul C. Canfield$^{\ast\dagger}$}
\affiliation{$^\ast$Ames Laboratory US DOE and $^\dagger$Department of Physics and Astronomy, Iowa State
University, Ames, IA 50011, USA}

\date{\today}

\begin{abstract}
Temperature- and field-dependent Hall effect measurements are reported for YbAgGe, a heavy fermion compound
exhibiting a field-induced quantum phase transition, and for two other closely related members of the RAgGe
series: a non-magnetic analogue, LuAgGe and a representative, ''good local moment'', magnetic material, TmAgGe.
Whereas the temperature dependent Hall coefficient of YbAgGe shows behavior similar to what has been observed in a
number of heavy fermion compounds, the low temperature, field-dependent measurements reveal well defined, sudden
changes with applied field; in specific for $H \perp c$ a clear local maximum that sharpens as temperature is
reduced below 2 K and that approaches a value of 45 kOe - a value that has been proposed as the $T = 0$ quantum
critical point. Similar behavior was observed for $H \| c$ where a clear minimum in the field-dependent Hall
resistivity was observed at low temperatures. Although at our base temperatures it is difficult to distinguish
between the field-dependent behavior predicted for (i) diffraction off a critical spin density wave or (ii)
breakdown in the composite nature of the heavy electron, for both field directions there is a distinct temperature
dependence of a feature that can clearly be associated with a field-induced quantum critical point at $T = 0$
persisting up to at least 2 K.
\end{abstract}


\maketitle

\section{Introduction}
Based on low temperature resistivity and heat capacity measurements in applied magnetic fields YbAgGe was recently
classified as a new heavy fermion material with long range, possibly small moment,  magnetic order below 1 K
\cite{mor04a,bud04a,kat04a,ume04a} that shows magnetic field induced non-Fermi-liquid (NFL) behavior
\cite{bud04a}. The critical field required to drive YbAgGe to the field-induced quantum critical point (QCP) is
anisotropic ($H_c^{ab} \approx$ 45 kOe, $H_c^c \approx$ 80 kOe) and conveniently accessible by many experimental
groups \cite{bud04a}. YbAgGe is one of the {\it rarae aves} of intermetallics (apparently only second, after the
extensively studied YbRh$_2$Si$_2$ \cite{tro00a,geg02a,ish02a,cus03a}) a {\it stoichiometric}, Yb - based, heavy
fermion (HF) that shows magnetic field induced NFL behavior and as such is suitable to serve as a testing ground
for experimental and theoretical constructions relevant for QCP physics. Among the surfeit of detailed
descriptions developed for a material near the antiferromagnetic QCP we will refer to the outcomes \cite{col01a}
of two more general, competing, pictures: in one viewpoint the QCP is a spin density wave (SDW) instability
\cite{ove59a} of the Fermi surface; within the second picture that originates in the description of heavy fermions
as a Kondo lattice of local moments \cite{don77a,don77b}, heavy electrons are composite bound states formed
between local moments and conduction electrons and the QCP is associated with the breakdown of this composite
nature. It was suggested \cite{col01a} that Hall effect measurements can help distinguish which of these two
mechanisms may be relevant for a particular material near a QCP. In the SDW scenario the Hall coefficient is
expected to vary continuously through the quantum phase transition, whereas in the composite HF scenario the Hall
coefficient is anticipated to change discontinuously at the QCP. Perhaps more importantly, in both scenarios a
clear and sharp change in the field dependent Hall effect (for the field-induced QCP) is anticipated to occur at
low temperatures, near the critical field value.

Although Hall effect measurements appear to be a very attractive method of gaining insight into the nature of the
QCP, one has to keep in mind that an understanding of the different contributions to the measured Hall
coefficient, in particular in magnetic or strongly correlated materials, is almost inevitably difficult and
potentially evasive \cite{hur72a,pas03a}. Therefore measurements on samples well characterized by other techniques
\cite{mor04a,bud04a} as well as comparison with non-magnetic as well as non-HF members of the same series can be
beneficial. In this work we present temperature- and field- dependent Hall effect measurements on YbAgGe single
crystals. The non-magnetic member of the same RAgGe (R = rare earth) series, LuAgGe, and the magnetic, essentially
non-hybridizing, TmAgGe were used for ''common sense'' checks, or calipers, of the YbAgGe measurements.

\section{Experimental}

YbAgGe, LuAgGe and TmAgGe single crystals in the form of clean, hexagonal-cross-section rods of several mm length
and up to 1 mm$^2$ cross section were grown from high temperature ternary solutions rich in Ag and Ge (see
\cite{mor04a} for details of the samples' growth). Their structure and the absence of impurity phases were
confirmed by powder X-ray diffraction. Temperature and field dependent Hall resistivity, $\rho_H(H,T)$, and
auxiliary high field magnetization measurements were performed down to 1.9 K, in an applied magnetic field of up
to 140 kOe in a Quantum Design PPMS-14 instrument. For YbAgGe Hall measurements were extended down to 0.4 K using
the He-3 option of the PPMS-14. A four probe, ac technique ($f$ = 16 Hz, $I$ = 1-0.1 mA), was used for the Hall
measurements. Samples were polished down to a plate-like shape with thicknesses of 0.3-0.4 mm. Pt leads were
attached to the sample with Epotek H20E silver epoxy so that the current was flowing along the crystallographic
$c$ axis. For the $H \| ab$ case Hall resistivity ($\rho_H$) was measured in the hexagonal crystallographic plane
(approximately along $[100]$ direction) with the magnetic field applied perpendicular to both the current and the
Hall voltage directions (approximately along $[120]$ direction) (see the lower inset to Figure \ref{rHLuTm}). In
the $H \| c$ case, current was flowing in the hexagonal plane, approximately in $[100]$ direction, the Hall
voltage was measured along $[120]$ direction. Due to rod-like morphology of the crystals, samples that were cut
and polished for $H \| c$ measurements were smaller and the error bars in the absolute values (due to geometry and
position of the contacts) are larger than for the $H \| ab$ data sets. To eliminate the effect of inevitable
(small) misalignment of the voltage contacts, the Hall measurements were taken for two opposite directions of the
applied field, $H$ and $-H$, and the odd component, $(\rho_H(H)-\rho_H(-H))/2$ was taken as the Hall resistivity.
To determine the Hall resistivity in the limit of low field, linear fits of the initial (linear) parts of the
$\rho(H)$ data in both quadrants were used. He-4 Hall measurements for YbAgGe and LuAgGe were performed on two
samples of each material, the results were the same within the error bars in sample dimensions and contact
position measurements. During the measurements particular care was taken to avoid rotation and/or misplacement of
the TmAgGe sample due to its magnetic anisotropy.

\section{Results and Discussion}

\subsection{LuAgGe and TmAgGe}
\begin{figure}
\begin{center}
\includegraphics[angle=0,width=100mm]{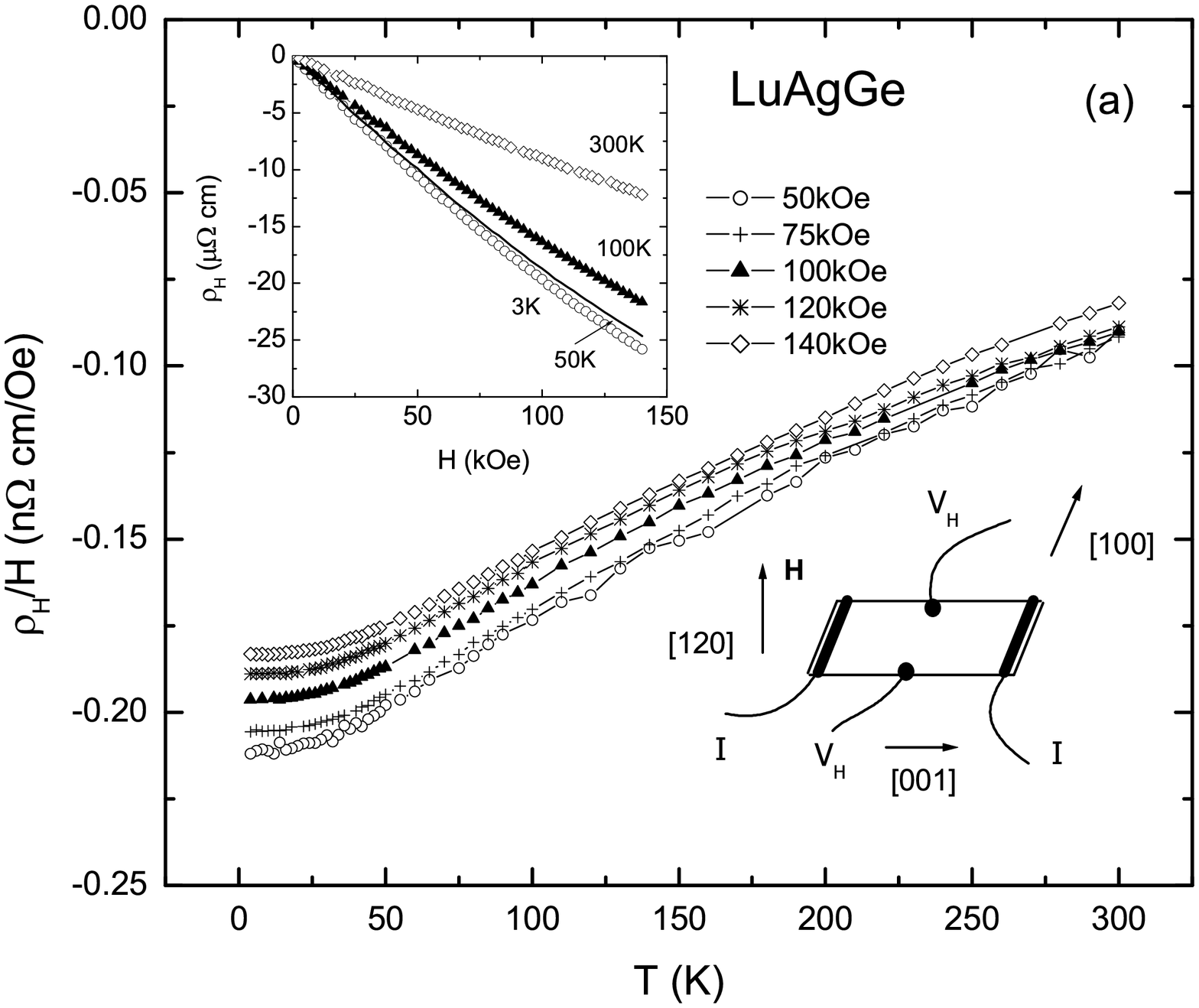}
\includegraphics[angle=0,width=100mm]{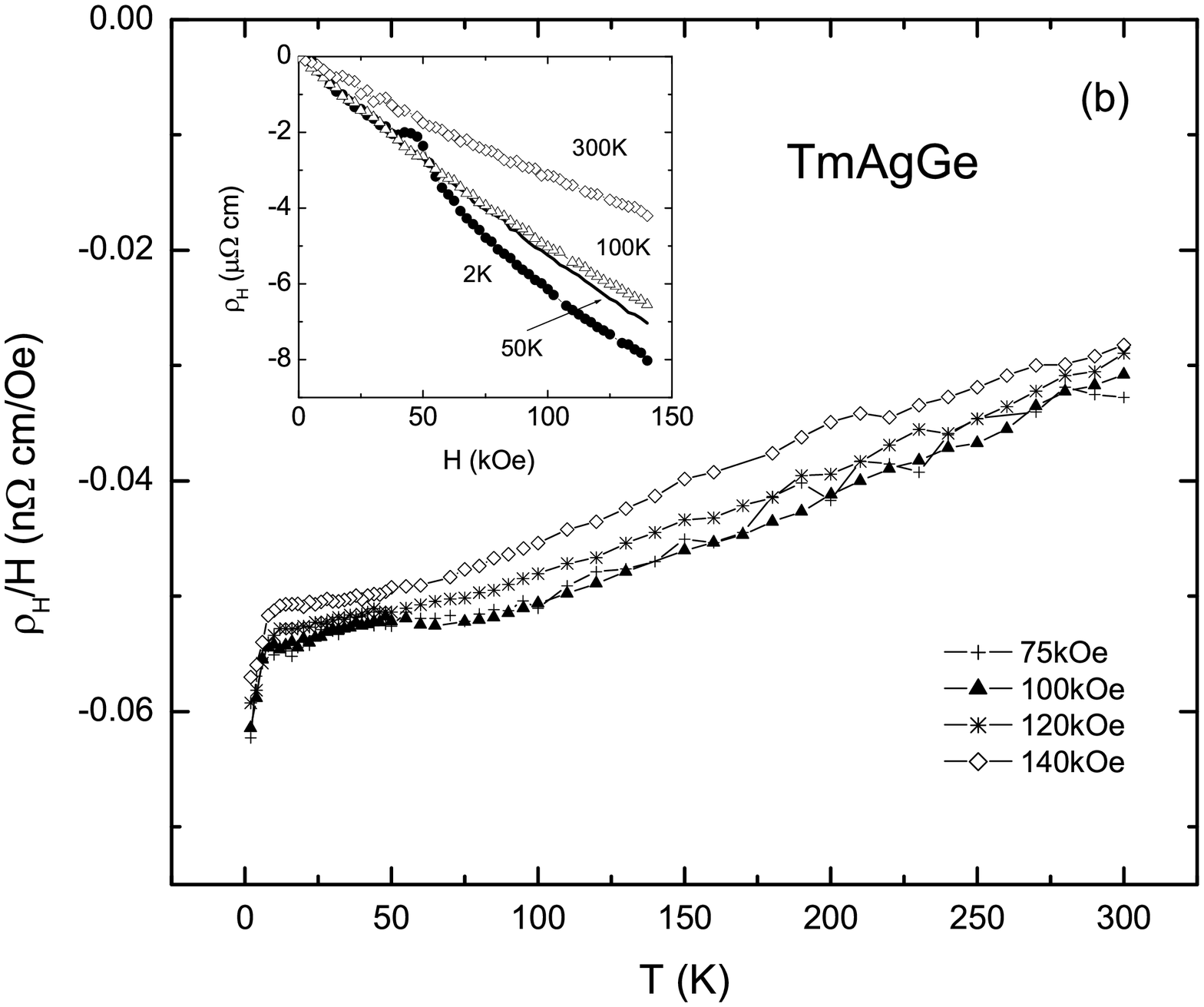}
\end{center}
\caption{(a) Temperature-dependent Hall coefficient, $\rho_H/H$, of LuAgGe measured in different applied fields
($H \| ab$). Upper inset: field-dependent Hall resistivity of LuAgGe measured at different temperatures. Lower
inset: the sample, current and applied field geometry used during the measurements. (b) Similar data for
TmAgGe.}\label{rHLuTm}
\end{figure}

The field-dependent Hall resistivity for LuAgGe for $H \| ab$ is shown in the upper inset to Fig. \ref{rHLuTm}(a)
for several temperatures. $\rho_H$ is only slightly non-linear in field over the whole temperature range. This
minor non-linearity causes some difference in the $\rho_H/H$ {\it vs.} $T$ data obtained in different applied
fields (Fig. \ref{rHLuTm}(a)). The Hall coefficient, $R_H = \rho_H/H$, is measured to be negative. The overall
temperature dependence is monotonic, slow and featureless with approximately a factor of two increase in the
absolute value of $\rho_H/H$ from room temperature to low temperatures. This temperature-dependency of the Hall
coefficient of the non-magnetic material possibly reflects some details of its electronic structure (for example,
comparable factor of 2 changes in $R_H$ were recently observed in LaTIn$_5$, T = Rh, Ir, Co, \cite{hun04a}).
Overall the temperature- and field-dependence of the Hall coefficient for TmAgGe (Fig. \ref{rHLuTm}(b)) is similar
to that of LuAgGe with two main differences: (i)the long range order and metamagnetism of TmAgGe \cite{mor04a} is
reflected in Hall measurements as a low temperature decrease in $R_H(T)$ and as anomalies in $\rho_H(H)$ for $T =$
2 K that are consistent with the fields of the metamagnetic transitions; (ii)the absolute values of the $R(H)$
data for TmAgGe are a factor of 3-4 smaller than for LuAgGe.

\subsection{YbAgGe, $H \| ab$}

\begin{figure}
\begin{center}
\includegraphics[angle=0,width=120mm]{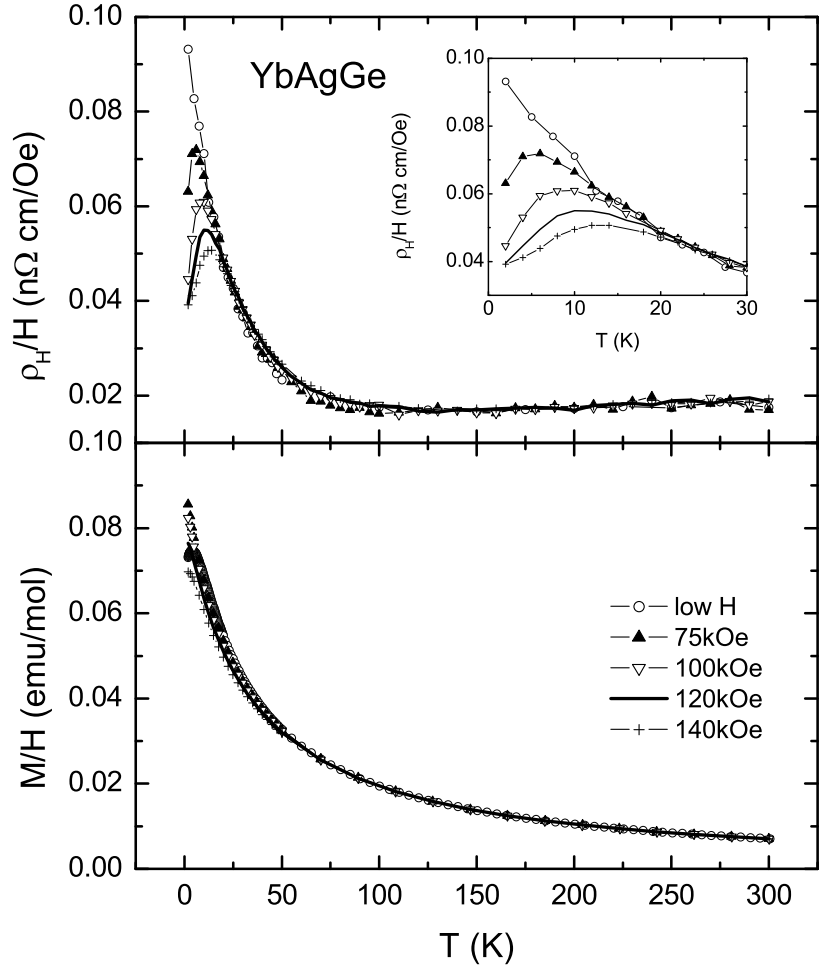}
\end{center}
\caption{Upper panel: temperature-dependent Hall coefficient, $\rho_H/H$, of YbAgGe measured in different applied
fields ($H \| ab$). Inset: enlarged, low temperature part of the data. Lower panel: DC susceptibility of YbAgGe
($H$ along $[120]$ direction). The ''low H'' label in the legend refers to the low field Hall resistivity (see
Experimental section) and for susceptibility measured in $H$ = 1 kOe.}\label{rHMTYb}
\end{figure}

\begin{figure}
\begin{center}
\includegraphics[angle=0,width=120mm]{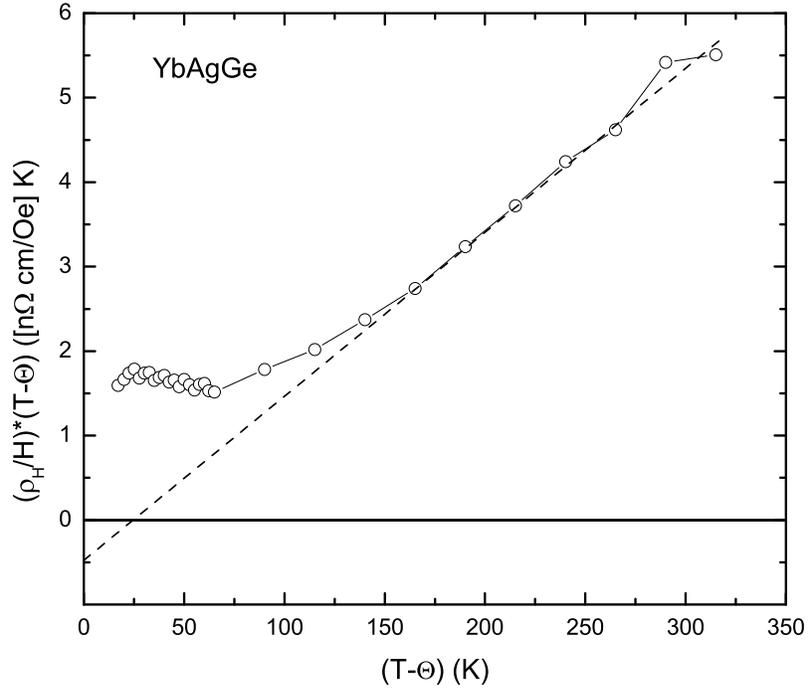}
\end{center}
\caption{Temperature-dependent low field Hall coefficient ($H \| ab$) plotted as $(\rho_H/H) \times (T-\Theta)$
{\it vs.} $(T-\Theta)$.}\label{RHT}
\end{figure}

The temperature dependent Hall coefficient and the DC susceptibility data for YbAgGe with the same orientation of
the magnetic field with respect to the crystallographic axis are shown in Fig. \ref{rHMTYb}. The susceptibility,
$M/H$, is field-independent above ~50 K ({\it i.e.} $M(H)$ is linear below 140 kOe in this temperature range) and
is similar to the data reported in \cite{mor04a,bud04a}. The Hall coefficient, $R_H$, is field-independent above
approximately 25 K. The temperature dependencies of the susceptibility and the Hall coefficient at high
temperatures closely resemble each other. At low temperatures a field-dependent maximum in $R_H$ (see inset to
Fig. \ref{rHMTYb}) is observed. Qualitatively the temperature dependence of the Hall coefficient is consistent
with the picture presented in \cite{col85a,had86a,fer87a,lap87a} (see also \cite{hur72a,chi80a} for a
comprehensive review). Within this picture the temperature dependence of the Hall coefficient in heavy fermion
materials is a result of two contributions: a residual Hall coefficient, $R_H^{res} = \rho_H^{res}/H$, and a Hall
coefficient due to the intrinsic skew scattering, $R_H^s = \rho_H^s/H$. The residual Hall coefficient is ascribed
to a combination of the ordinary Hall effect and residual skew scattering by defects and impurities and, to the
first approximation, is considered to be temperature-independent, although, realistically, both the ordinary Hall
effect and the residual skew scattering may have weak temperature dependence. The temperature-dependent, intrinsic
skew scattering contribution ($R_H^s$) at high temperatures ($T \gg T_K$, where $T_K$ is the Kondo temperature)
increases as the temperature is lowered in a manner that is mainly due to the increasing magnetic susceptibility.
At lower temperatures $R_H^s$ passes through a crossover regime, then has a peak at a temperature on the order of
the coherence temperature, $T_{coh}$, and finally, on further cooling rapidly decreases (in the coherent regime of
skew scattering by fluctuations) to zero ({\it i.e.} $R_H$ ultimately levels off to the $\sim R_H^{res}$ value at
very low temperatures \cite{had86a,fer87a,lap87a}). In the high temperature ($T \gg T_K$) limit we can (very
roughly, within an order of magnitude) separate these two contributions to the observed temperature-dependent Hall
coefficient using a phenomenological expression $R_H(T) = R_H^{res} + R_s \times \chi(T)$ \cite{oha80a} with the
temperature-dependent skew scattering contribution written as $R_H^s(T) = R_s \times \chi(T)$ where $\chi(T) =
C/(T - \Theta)$, $C$ is the Curie constant, and $\Theta$ is the Weiss temperature. Using $\Theta_{ab}$ = -15.1 K
from \cite{mor04a} we can plot $R_H(T) \times (T-\Theta)$ {\it vs.} $(T-\Theta)$ (Fig. \ref{RHT}) and from the
linear part of the curve we can estimate $R_H^{res} \approx$ 0.02 n$\Omega$ cm/Oe and $R_s \approx$ -0.17
n$\Omega$ cm/Oe. It seems peculiar that our estimate of $R_H^{res}$ for YbAgGe differs noticeably from the Hall
coefficient measurements for LuAgGe and TmAgGe (see Fig. \ref{rHLuTm}). Regarding this discrepancy it should be
mentioned that besides possible experimental (mainly geometrical) errors these three materials may have different
residual skew scattering and, additionally, as indicated by the preliminary results of band structure calculations
\cite{sam04a}, the density of states at the Fermi level can be considerably different for all three compounds
under consideration. Although the magnetic susceptibility, $\chi(T)$ of TmAgGe above the N\'{e}el temperature has
a clear Curie - Weiss behavior \cite{mor04a}, in contrast to the case of YbAgGe, the temperature dependence of the
Hall coefficient for TmAgGe (Fig. \ref{rHLuTm}(b)) does not have a similar functional form. The reason for this
difference is apparently the very small skew scattering contribution ($R_s \ll R_H^{res}$) to the Hall coefficient
in TmAgGe. Similarly small couplings of local moment magnetism with the Hall effect has been seen in other rare
earth intermetallics, {\it e.g.} RNi$_2$B$_2$C (R = rare earth) borocarbides \cite{fis95a,man97a,nar99a}.

\begin{figure}
\begin{center}
\includegraphics[angle=0,width=120mm]{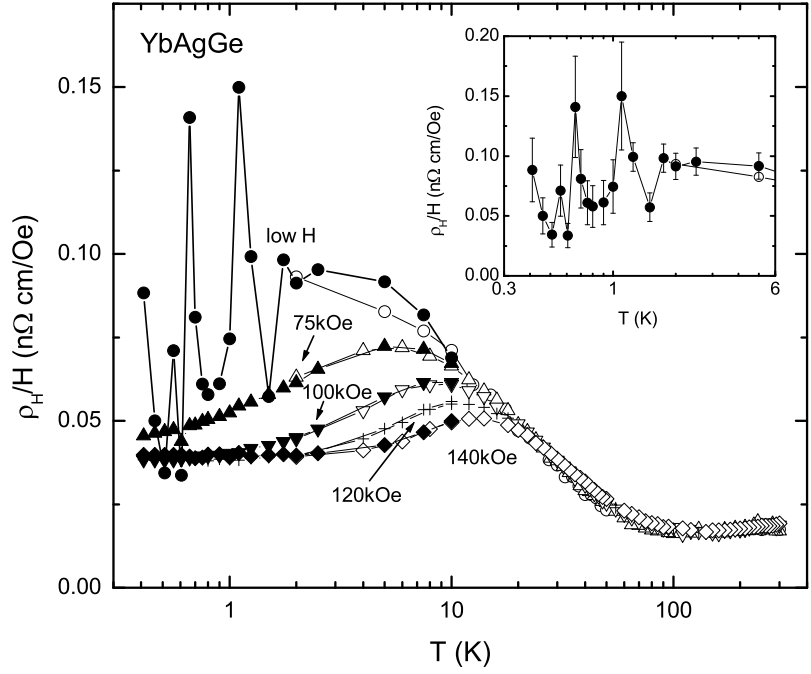}
\end{center}
\caption{Temperature-dependent Hall coefficient of YbAgGe ($H \| ab$) measured in different applied fields down to
0.4 K. Open symbols - He-4 measurements (2-300 K), filled symbols - measurements using He-3 option (0.4-10 K).
Inset: enlarged low temperature part of the low field data with the estimated error bars.}\label{rHLTYb}
\end{figure}

In order to further explore the low temperature behavior of the Hall coefficient, measurements down to 0.4 K were
performed. The results (on a semi -{\it log} scale) are shown in Fig. \ref{rHLTYb}. The data taken in applied
fields of 75 kOe and higher show the expected levelling off of the $R_H(T)$ as $T \to 0$. It is noteworthy that
the measured value of $R_H(T \to 0)$ is close to the aforementioned estimate of the residual Hall coefficient.
This agreement suggests that at the lowest temperatures the Hall coefficient is dominated by $R_H^{res}$ and,
barring the residual skew scattering contribution, can probe the concentration of the electronic carriers.

Whereas the higher field values of the Hall coefficient vary smoothly with temperature (Fig. \ref{rHLTYb}), the
low field data, below $T \approx$ 3 K, show large variations. Although the signal to noise ratio in the low field
measurements is inherently lower, these variations appear to be above the noise level (Fig. \ref{rHLTYb}, inset)
and the peaks slightly above 0.6 K and 1.0 K are understood as the signatures of the magnetic transitions in
YbAgGe \cite{mor04a,bud04a,ume04a} that are suppressed (in this orientation) when a 75 kOe, or higher, magnetic
field is applied.

To further study the field-induced QCP in YbAgGe, field dependent Hall resistance measurements were performed at
different temperatures (Fig. \ref{rHHYb}). Although the theoretical constructions are usually formulated in terms
of the Hall {\it coefficient}, not Hall {\it resistivity}, in the case of YbAgGe the magnetic field itself is a
control parameter for the QCP that makes the proper definition of the Hall coefficient ambiguous. We will continue
presenting our data as Hall resistivity, since it is a quantity unambiguously extracted from the measurements, and
leave the discussion on the suitable definition of the Hall coefficient for the Appendix.

For temperatures at and above $\sim$10 K, the $\rho_H(H)$ behavior is monotonic and, at higher temperatures,
eventually linear (Fig. \ref{rHHYb}(b)). This type of behavior has been observed in a number of different
materials in the paramagnetic state \cite{hur72a}. The low temperature evolution of the $\rho_H(H)$ behavior is
more curious (Fig. \ref{rHHYb}(a)) and ought to be compared with the phase diagram obtained for YbAgGe ($H \| ab$)
in \cite{bud04a} (an augmented version of which is shown in Fig. \ref{PD} below). The lines in Fig. \ref{rHHYb}(a)
roughly connect the points according to the phase lines in \cite{bud04a} (see also Fig. \ref{PD} below). It can be
seen that the lower $H-T$ magnetically ordered phase line possibly has (despite the scattering of the points)
correspondent features in $\rho_H(H)$, and the coherence line in \cite{bud04a} (and Fig. \ref{PD}) roughly
corresponds to the beginning of the high field linear behavior in $\rho_H(H)$. On the other hand, the higher $H-T$
magnetically ordered phase line cannot unambiguously be associated with any feature in $\rho_H(H)$ curves.

The most interesting feature shown in Fig. \ref{rHHYb}(a) though is the presence of the pronounced peak, or local
maximum, in $\rho_H(H)$ that occurs at $\approx 45$ kOe for the $T =$ 0.4 K curve and can be followed up to
temperatures above long range magnetic order transition temperatures. For $T =$ 2.5 K a broad, local maximum in
$\rho_H$, centered at $H \approx 100$ kOe can just barely be discerned. As temperature is reduced this feature
sharpens and moves down in field. For $T =$ 1 K the local maximum in $\rho_H$ is clearly located at $H \approx 50$
kOe and by $T = 0.4$ K $\rho_H$ has sharpened almost to the point of becoming discontinuous with $H_{max} \approx
45$ kOe. The temperature dependence of $H_{max}$ is shown in Fig. \ref{PD} clearly demonstrating that as $T \to
0$, $H_{max} \to H_{crit}$ for the QCP. Independent of any theory these data clearly show that (i) $\rho_H$ is an
extremely sensitive method of determining $H_{crit}$ of QCP, (ii) $H_{max}$ has a clear temperature dependence,
and (iii) the QCP influences $\rho_H$ up to $T \leq 2.5$ K, a temperature significantly higher than the $H = 0$
antiferromagnetic ordering temperature.

The new phase line (shown as stars in Fig. \ref{PD}) associated with $\rho_H$ maximum is distinct from the lines
inferred from $C_p(T,H)$ and $\rho(T,H)$ data \cite{bud04a}. As $T \to 0$ this line approaches $H_{crit}$, but for
finite $T$ it is well separated from the coherence line that was determined by the onset of $T^2$ resistivity
behavior. This new $H_{max}$ line rather clearly locates $H_{crit}$ at $\sim 45$ kOe, the field at which the long
range antiferromagnetic order appears to be suppressed.

\subsection{YbAgGe, $H \| c$}

\begin{figure}
\begin{center}
\includegraphics[angle=0,width=120mm]{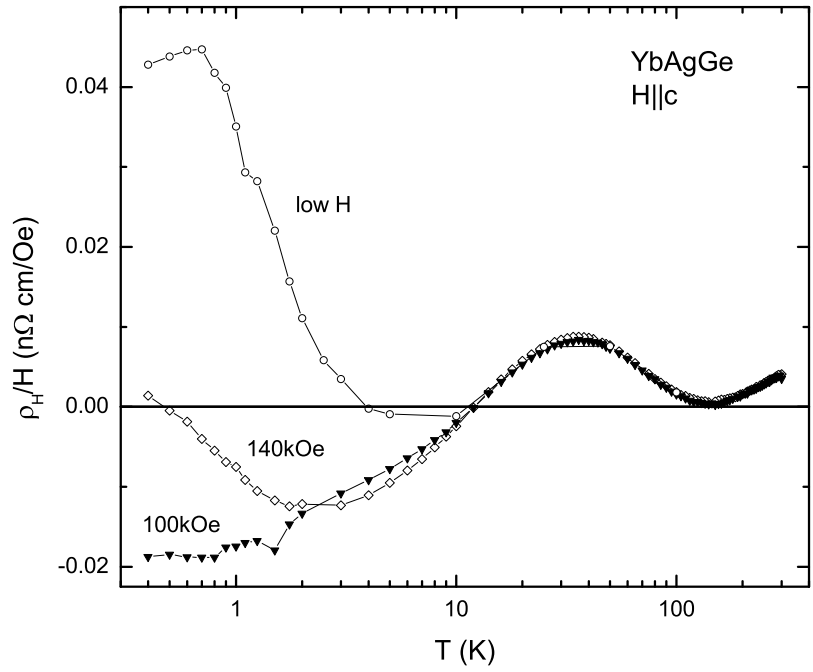}
\end{center}
\caption{Temperature-dependent Hall coefficient of YbAgGe ($H \| c$) measured in different applied fields down to
0.4 K.}\label{rHLTYbC}
\end{figure}

Since the response of YbAgGe to an applied magnetic field is anisotropic \cite{mor04a,bud04a,kat04a,ume04a}, it is
apposite to repeat the Hall measurements for the magnetic field applied parallel to the crystallographic $c$-axis.
The temperature-dependent Hall coefficient taken in different applied fields is presented in Fig. \ref{rHLTYbC}
(the low-field data were obtained as described above). The $R_H(T)$ behavior for $H \| c$ is qualitatively similar
to that for $H \| ab$ with a broad maximum being shifted to $\sim 30$ K (as compared to $\sim 10$ K for $H \| ab$)
and being less sensitive to the applied field. The low temperature, field-dependent Hall resistivity for $H \| c$
is shown in Fig. \ref{rHHYbC}. In many aspects the overall behavior is similar to that for $H \| ab$: there are no
apparent features associated with the phase lines derived from magnetoresistance and specific heat measurements
\cite {bud04a} (shown as lines in Fig. \ref{rHHYbC}), however there is the presence of a pronounced minimum in
$\rho_H(H)$ that occurs at $\approx 98$ kOe for the $T =$ 0.4 K curve and can be followed up to the temperatures
well above the zero-applied-field magnetic transition temperatures. For $T =$ 2 K a broad, local minimum in
$\rho_H$, centered at $H \approx 128$ kOe can still be recognized and at $T = 2.5$ K a local minima occurs just at
the edge of our field range. As temperature is reduced this feature sharpens and moves down in field. The
temperature dependence of $H_{min}$ is shown in Fig. \ref{PDC} clearly demonstrating that, akin to the $H \| ab$
case, as $T \to 0$, $H_{min} \to H_{crit}$ for the QCP.  The $\rho_H(H)$ behavior for this orientation is more
complex, and there is an additional, broad maximum in lower fields ($H \approx 50$ kOe at 0.4 K) that fades out
with increasing temperature. This highly non-monotonic in field behavior is the origin of the dissimilarities in
the low temperature $R_H(T)$ data (Fig. \ref{rHLTYbC}) taken in different applied fields.

The high field minimum in $\rho_H(H)$ (Fig. \ref{rHHYbC}) defines a new phase line (shown as stars in Fig.
\ref{PDC}) which is clearly different from the lines inferred from $C_p(T,H)$ and $\rho(T,H)$ data \cite{bud04a}.
As $T \to 0$ this line approaches $H_{crit}$, but for finite $T$ it is well separated from the coherence line that
was determined by the onset of $T^2$ resistivity behavior. For this orientation of the applied field this new
$H_{min}$ line rather clearly locates $H_{crit}$ at $\sim 100$ kOe, the field at which the long range
antiferromagnetic order appears to be suppressed.

It should be noted that the new lines in the $H-T$ phase diagrams were established from different types of extrema
in $\rho_H(H)$, {\it maximum} for $H \| ab$ and {\it minimum} for $H \| c$. We neither consider this difference as
a reason for particular discomfort nor do we necessarily view it as a potential clue to deeper understanding of
the nature of the field-induced QCP in this material. The preliminary band structure calculations \cite{sam04a} on
LuAgGe, the non-magnetic analogue of the title compound, suggest that the members of the RAgGe series have a
complex Fermi surface consisting of multiple sheets. In such a case a change in the Fermi surface may possibly
have different signatures in the Hall measurements with different field orientation. In addition, existing QCP
models appear not to be at the level of considering different shapes and topologies of the Fermi surfaces.

Whereas these new, $H_{max}/H_{min}$ lines on the $H - T$ phase diagrams (Figs. \ref{PD} and \ref{PDC}) appear to
be closely related with the QCP their detailed nature and temperature dependencies will require further
experimental and theoretical attention.

\section{Summary}

The temperature- and field-dependent Hall resistivity have been measured for YbAgGe single crystals with $H \| ab$
and $H \| c$ orientation of the applied magnetic field. The temperature dependent Hall coefficient of YbAgGe
behaves similarly to other heavy fermion materials. Low temperature, field-dependent measurements reveal a local
maximum ($H \| ab$) or minimum ($H \| c$) in $\rho_H(H)$ for $T \leq 2.5$ K that occurs at a value that approaches
$H_{crit} \approx 45$ kOe ($H \| ab$) and $H_{crit} \approx 90$ kOe ($H \| c$) as $T \to 0$. These data indicate
that (i) the Hall resistivity is indeed a useful measurement for the study of QCP physics and (ii) the influence
of the QCP extends to temperatures significantly higher than the $H = 0$ antiferromagnetic ordering temperature.

\section{Appendix}

\begin{figure}
\begin{center}
\includegraphics[angle=0,width=90mm]{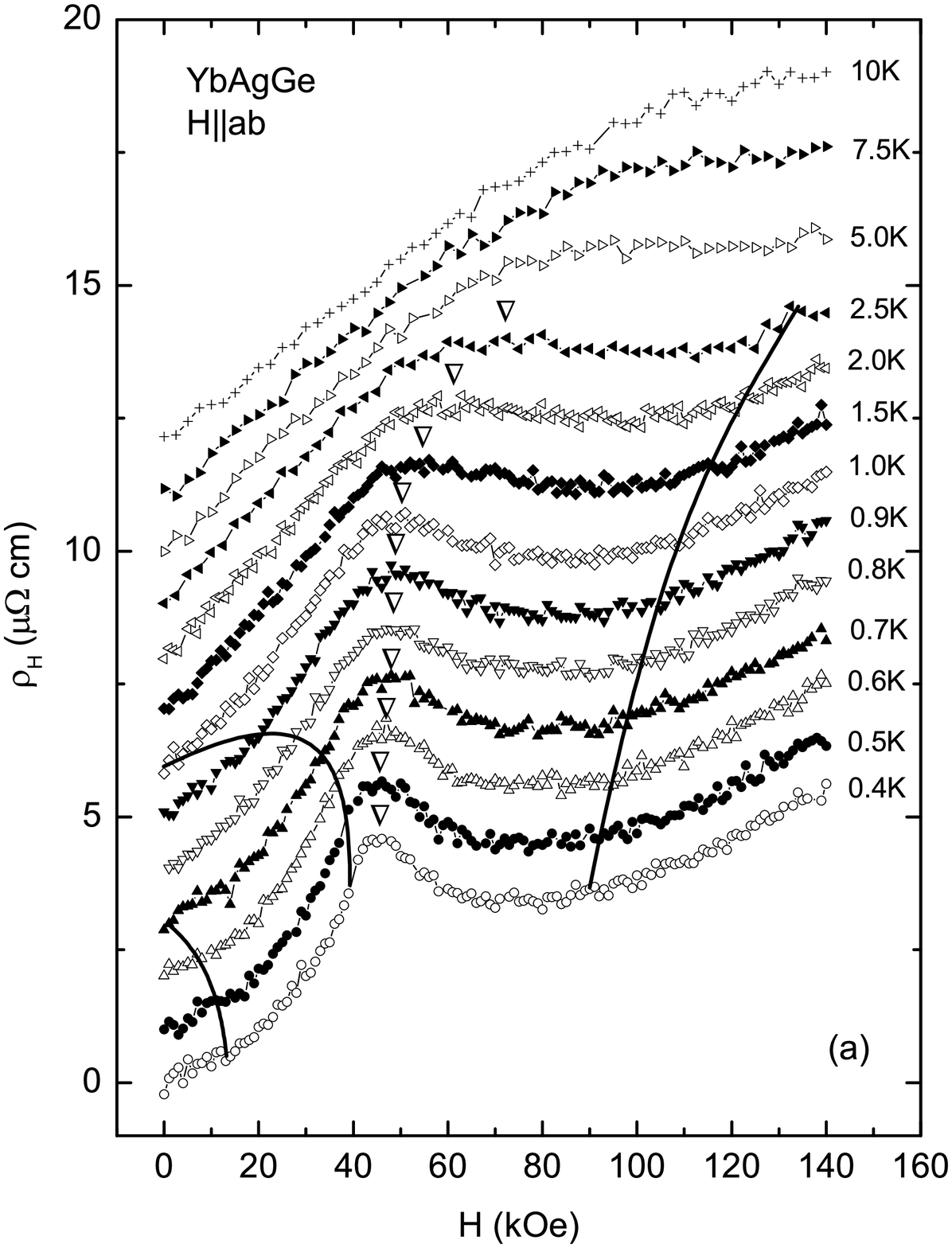}
\includegraphics[angle=0,width=90mm]{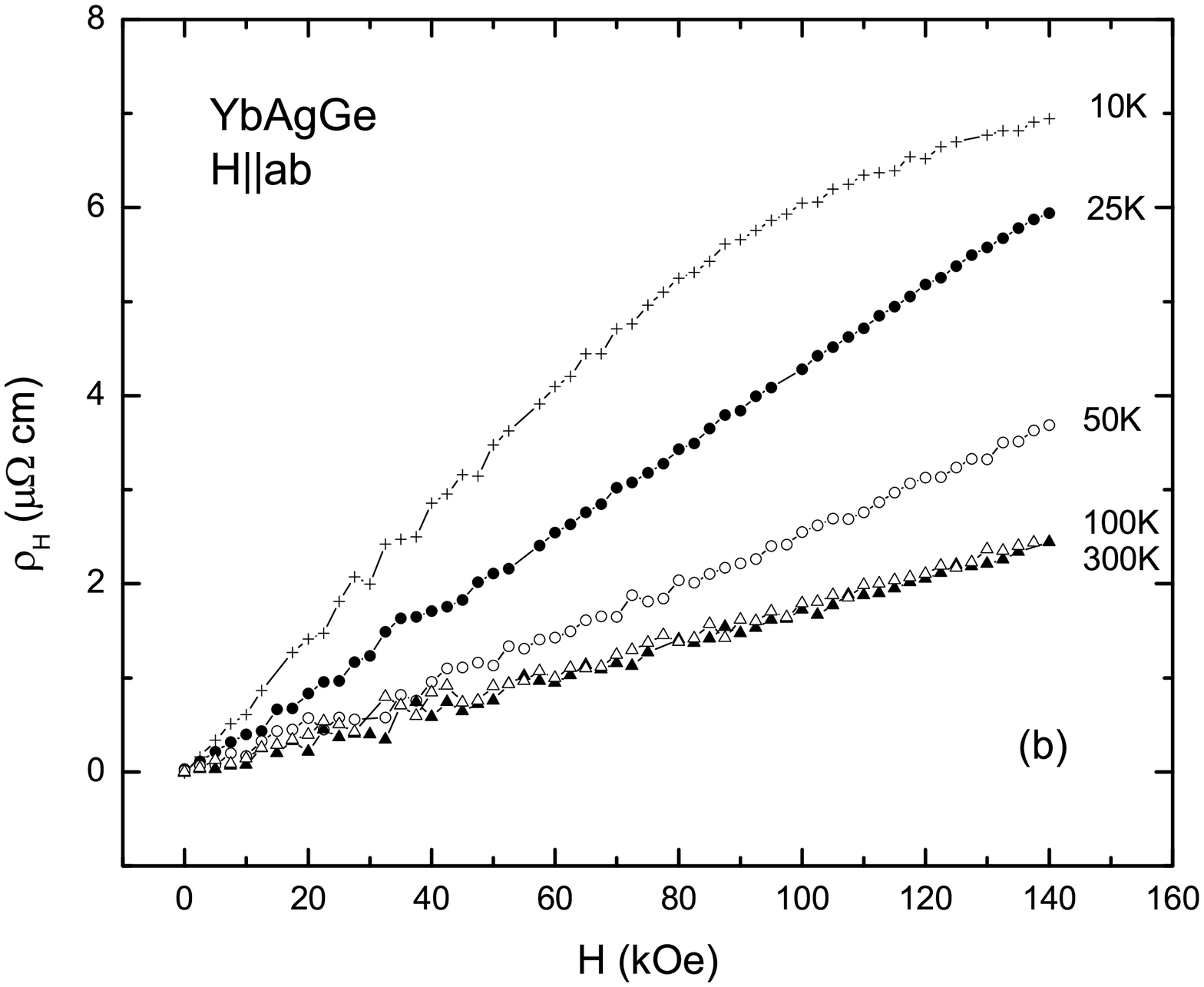}
\end{center}
\caption{Field-dependent Hall resistivity of YbAgGe ($H \| ab$) measured at different temperatures: (a)low
temperature data: the curves, except for $T$ = 0.4 K, are shifted by $1 \mu \Omega$ cm increments for clarity; the
lines represent the phase lines from the phase diagram in Fig. 10(a) of the Ref. 2; the triangles mark the
position of the local maximum in $\rho_H(H)$; (b)intermediate and high temperature data. Note: $T = 10$ K data is
shown in both plots for reference and is un-shifted in (b). }\label{rHHYb}
\end{figure}

\begin{figure}
\begin{center}
\includegraphics[angle=0,width=120mm]{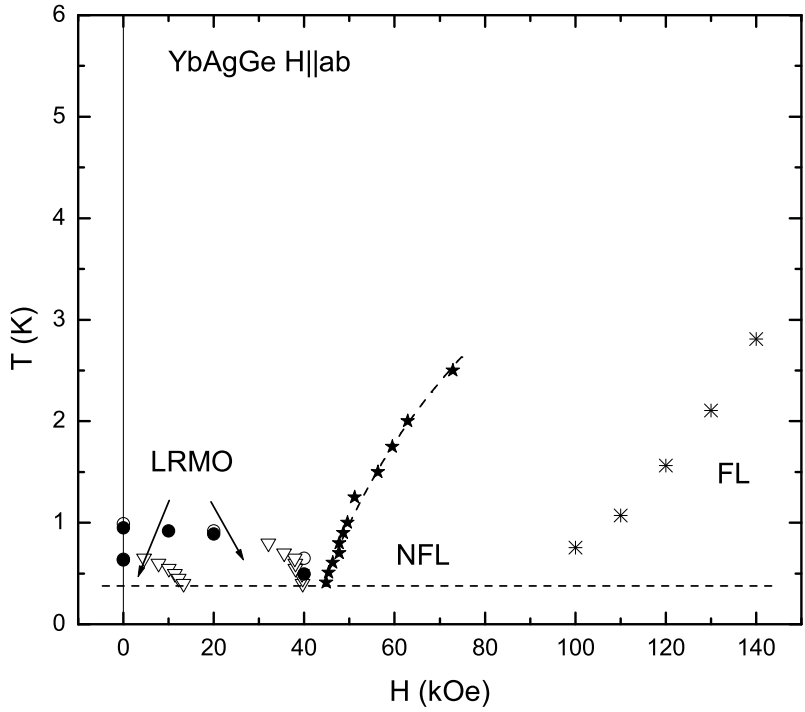}
\end{center}
\caption{Revised tentative $T - H$ phase diagram for $H
\parallel ab$. Long range magnetic order (LRMO) and the coherence temperature lines marked on the phase diagram
are taken from Ref. 2. Filled stars and corresponding dashed line as a guide to the eye are defined from the
maximum in the $\rho_H(H)$ curves.}\label{PD}
\end{figure}

\begin{figure}
\begin{center}
\includegraphics[angle=0,width=120mm]{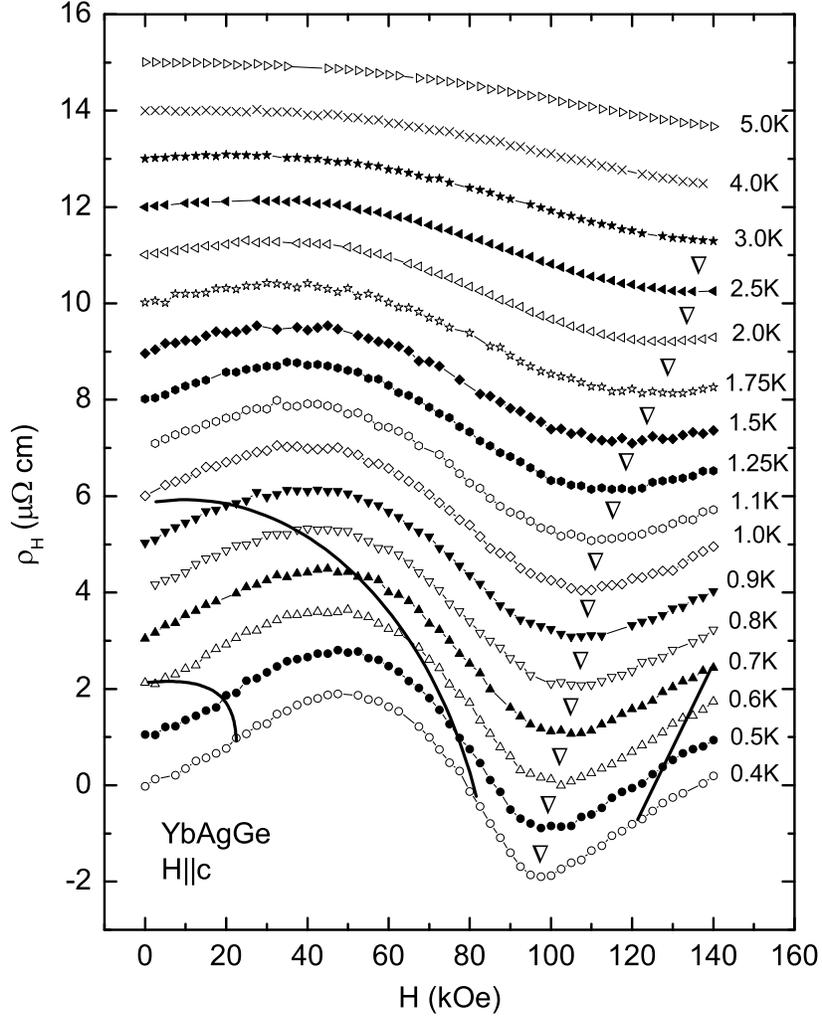}
\end{center}
\caption{Low temperature field-dependent Hall resistivity of YbAgGe ($H \| c$); the curves, except for $T$ = 0.4
K, are shifted by $1 \mu \Omega$ cm increments for clarity; the lines represent the phase lines from the phase
diagram in Fig. 10(b) of the Ref. 2; the triangles mark the position of the peak in $\rho_H(H)$. }\label{rHHYbC}
\end{figure}

\begin{figure}
\begin{center}
\includegraphics[angle=0,width=120mm]{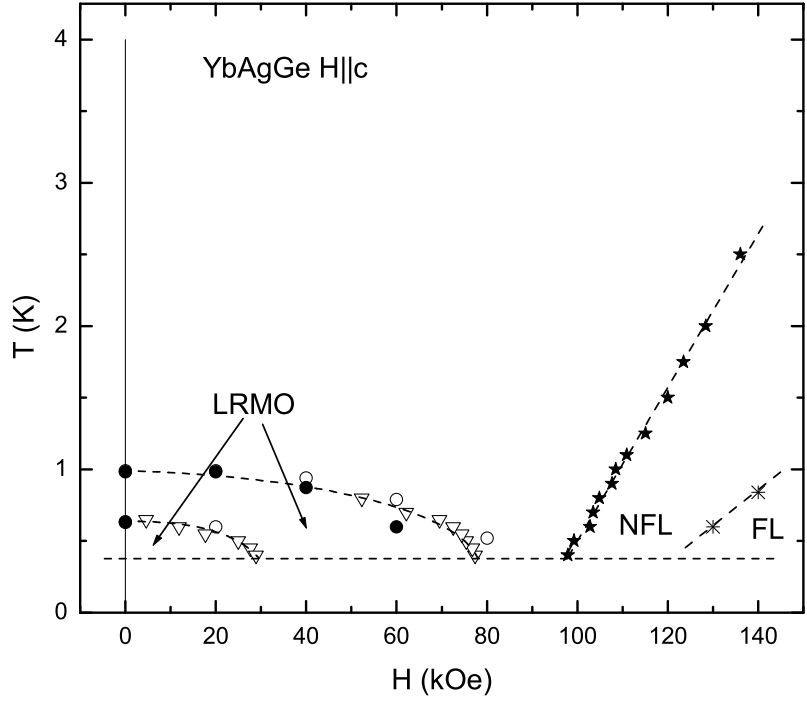}
\end{center}
\caption{Revised tentative $T - H$ phase diagram for $H
\parallel c$. Long range magnetic order (LRMO) and the coherence temperature lines marked on the phase diagram
are taken from Ref. 2. Filled stars are defined from the minimum in the $\rho_H(H)$ curves. Dashed lines are
guides to the eye.}\label{PDC}
\end{figure}

\begin{figure}
\begin{center}
\includegraphics[angle=0,width=75mm]{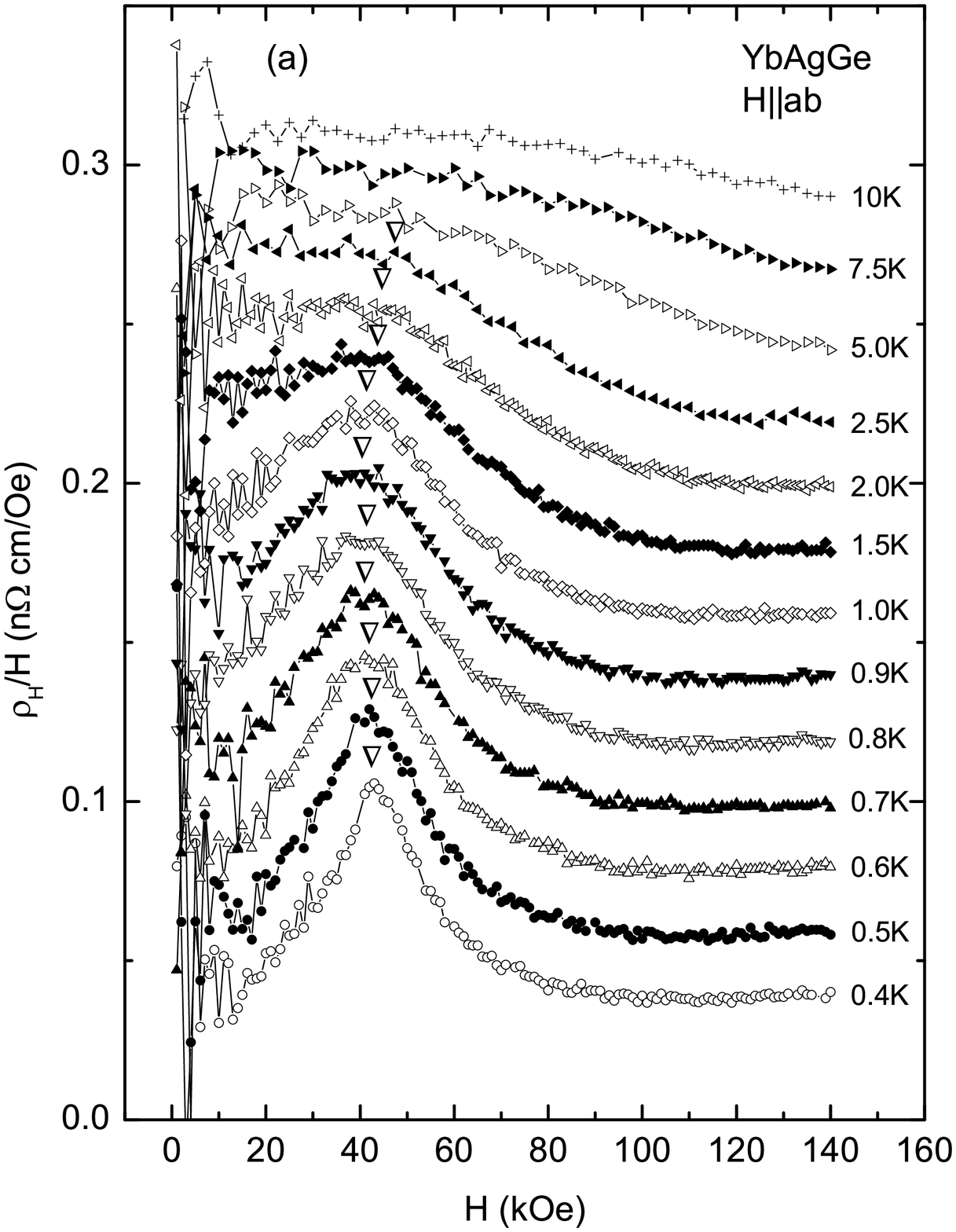}
\includegraphics[angle=0,width=75mm]{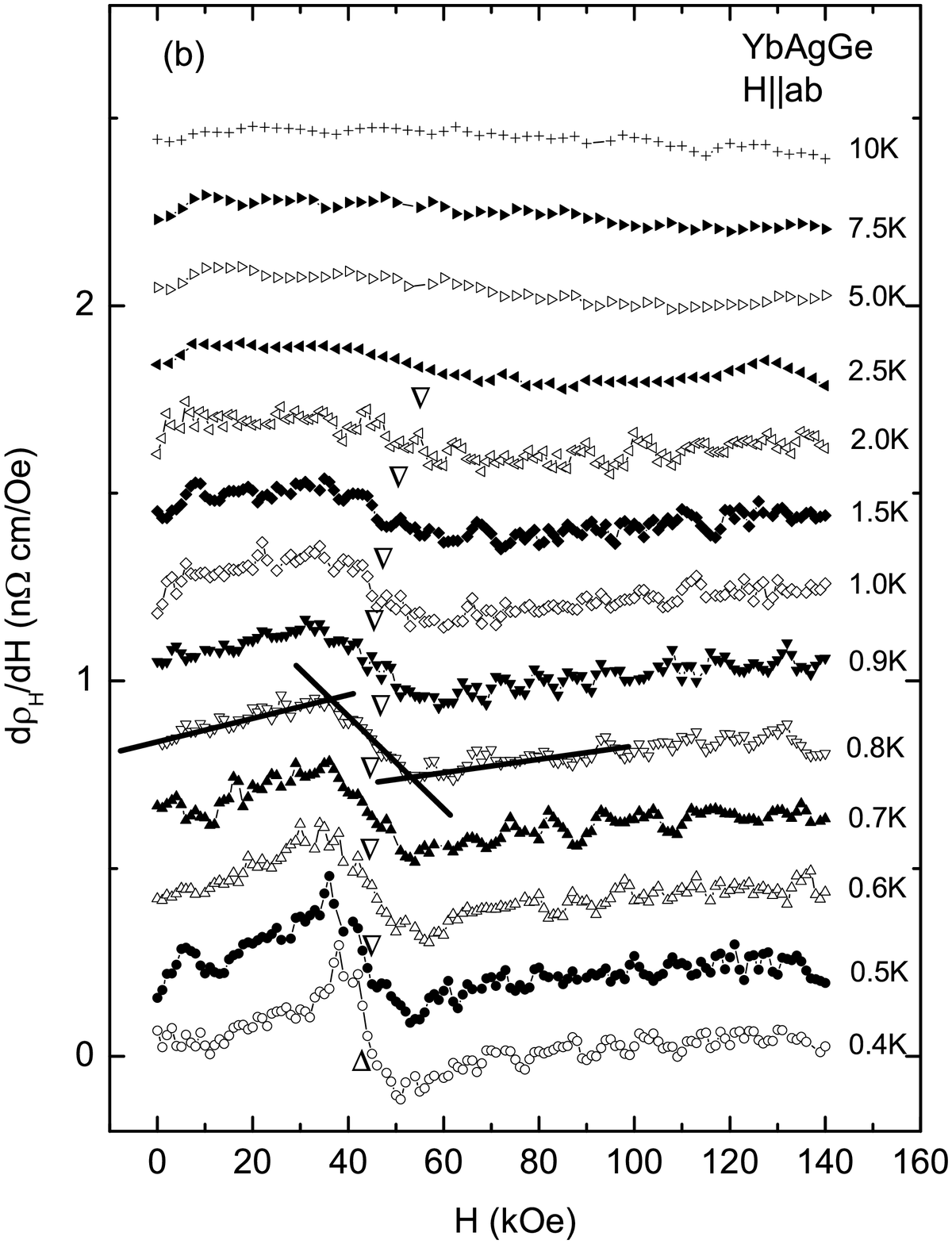}
\end{center}
\caption{Field-dependent Hall coefficient of YbAgGe ($H \| ab$), defined as (a) $R_H = \rho_H/H$ and (b) $R_H =
d\rho_H/dH$, measured at different temperatures. The curves, except for $T$ = 0.4 K, are shifted by (a) 0.02
n$\Omega$ cm and (b) 0.2 n$\Omega$ cm increments for clarity; the triangles mark the position of the feature in
$R_H(H)$: a local maximum in $\rho_H/H$ and a mid-point of the transition between two different field-dependent
regimes (see {\it e.g.} 0.8 K curve)in $d\rho_H/dH$. Curves in the (b) panel were obtained by differentiation of
the 5-adjacent-points-smoothed $\rho_H(H)$ data. Small downturn at $H \ge 130$ kOe in some $d\rho_H/dH$ curves
(panel (b)) is most likely an artifact of using digital smoothing and differentiation.}\label{drHHYb}
\end{figure}

\begin{figure}
\begin{center}
\includegraphics[angle=0,width=75mm]{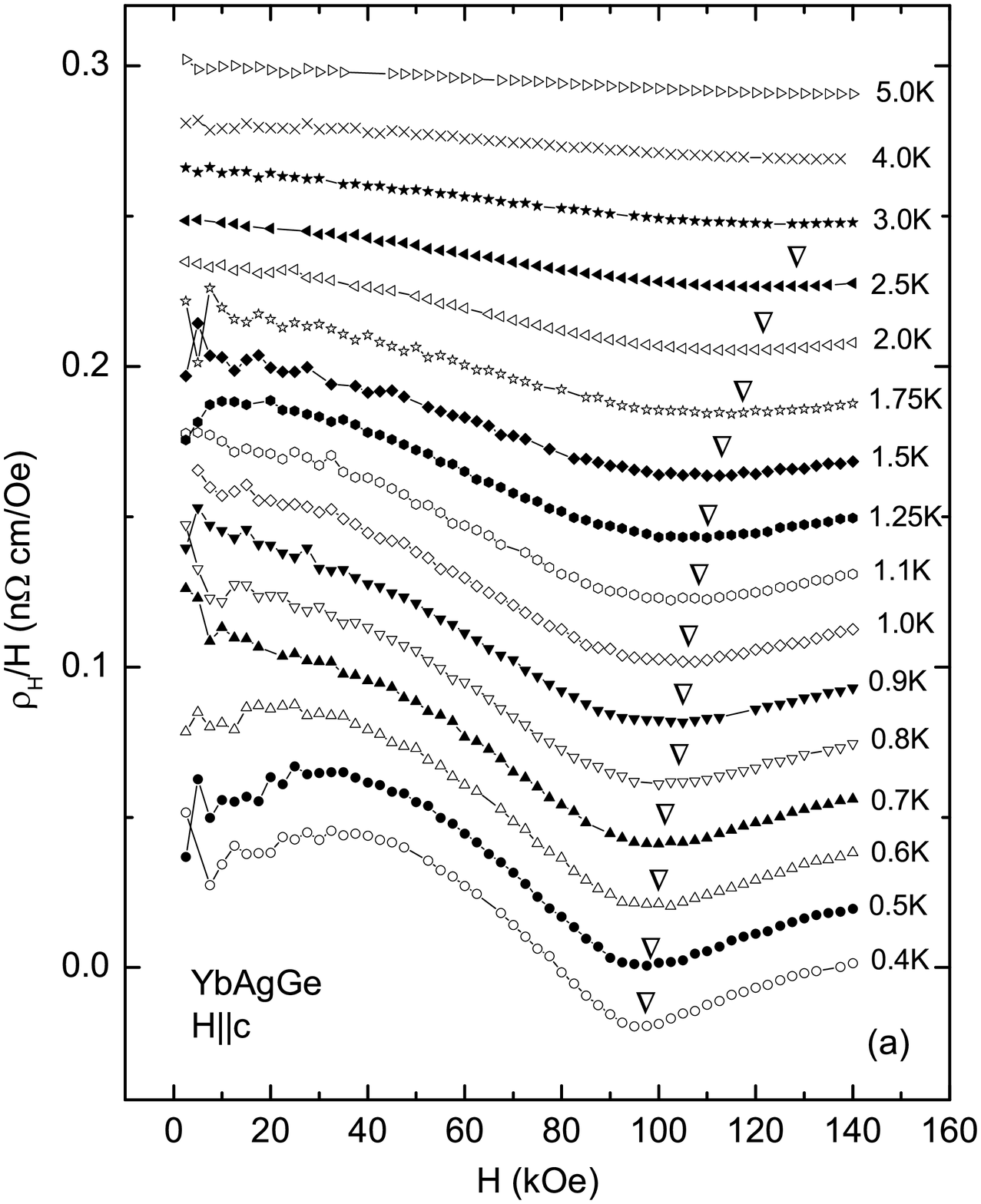}
\includegraphics[angle=0,width=75mm]{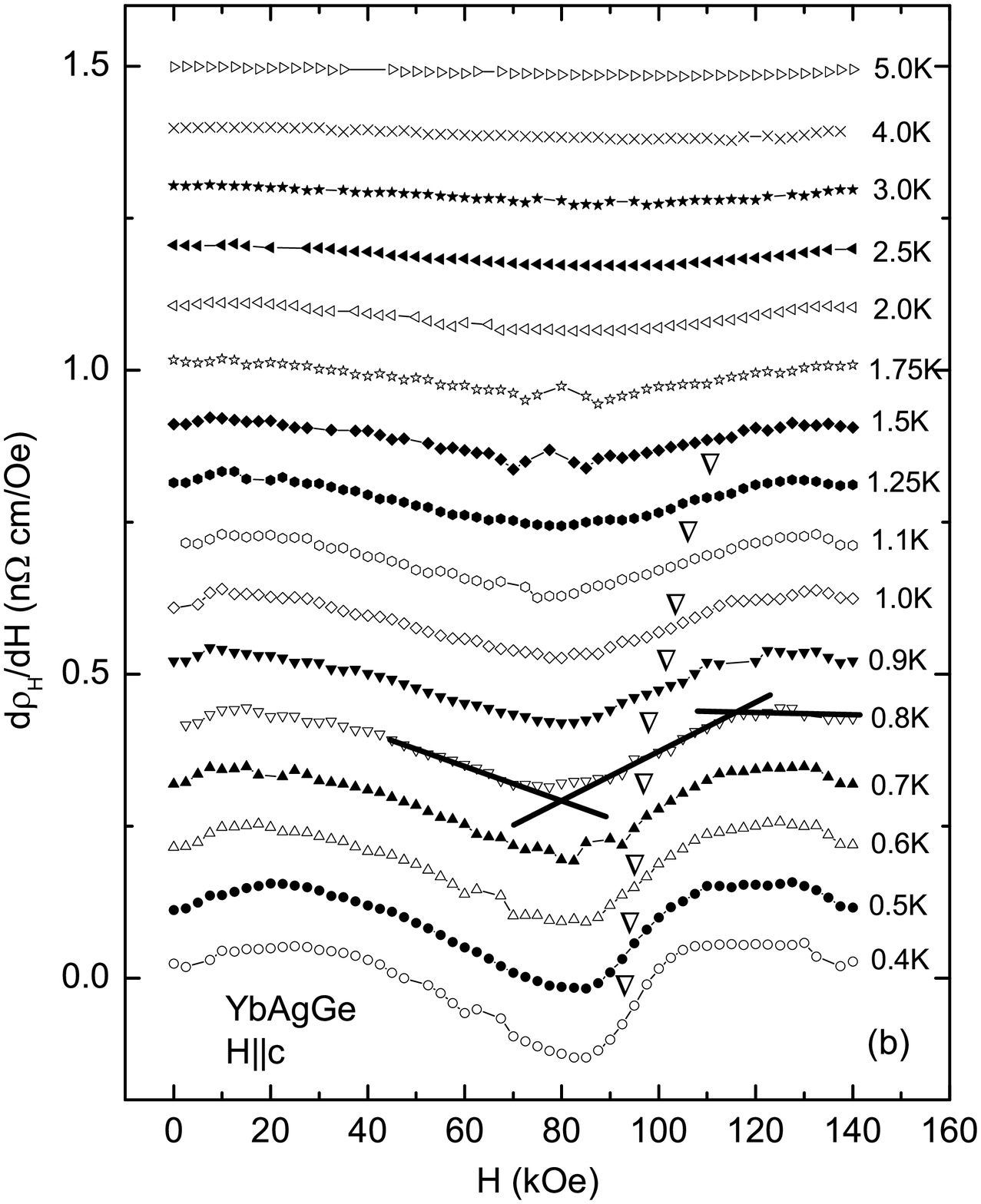}
\end{center}
\caption{Field-dependent Hall coefficient of YbAgGe ($H \| c$), defined as (a) $R_H = \rho_H/H$ and (b) $R_H =
d\rho_H/dH$, measured at different temperatures. The curves, except for $T$ = 0.4 K, are shifted by (a) 0.02
n$\Omega$ cm and (b) 0.1 n$\Omega$ cm increments for clarity; the triangles mark the position of the feature in
$R_H(H)$: a local minimum in $\rho_H/H$ and a mid-point of the transition between two different field-dependent
regimes (see {\it e.g.} 0.8 K curve)in $d\rho_H/dH$. Curves in the (b) panel were obtained by differentiation of
the 5-adjacent-points-smoothed $\rho_H(H)$ data. Small downturn at $H \ge 130$ kOe in some $d\rho_H/dH$ curves
(panel (b)) is most likely an artifact of using digital smoothing and differentiation.}\label{drHHYbC}
\end{figure}

Coleman {\it et al.} \cite{col01a} suggest that $R_H(P)$ data (where $P$ is a control parameter, {\it i.e.} $H$ in
our case) can be used to distinguish between two possible QCP scenarios: diffraction off of a critical spin
density wave or a breakdown of the composite nature of the heavy electron, with the former manifesting a change of
slope at $P_{crit}$ and the latter manifesting a divergence in the slope of $R_H(P)$ at $P_{crit}$. Since in our
case the magnetic field is itself the control parameter, it in not clear if $R_H = \rho_H/H$, $R_H = d\rho_H/dH$
or just simply $\rho_H$ should be used for comparison with the theory. $R_H(H)$ curves determined by two
aforementioned ways are presented in Fig. \ref{drHHYb} ($H \| ab$) and Fig. \ref{drHHYbC} ($H \| c$). For both
definitions and both orientations the evolution of a clear feature in $R_H(H)$ (defined as a local extremum for
$\rho_H/H$ and as a mid-point between two different field-dependent regimes for $d\rho_H/dH$)  replicates (albeit
with slight $H$-shift) the behavior of the Hall resistivity (Figs. \ref{rHHYb}(a), \ref{rHHYbC}). Given that the
new phase line in Figs. \ref{PD} and \ref{PDC} is fairly insensitive to the data analysis we feel that the use of
$\rho_H(H)$ data is currently the least ambiguous data set to analyze. On the other hand, if the form of the
anomaly near $H_{crit}$ is to be analyzed in detail it will be vital to have a more detailed theoretical treatment
of magneto-transport in field-induced QCP materials.

It is tempting to say that for the case of applied field as a control parameter the quantity $d\rho_H(H)/dH$
(rather than $\rho_H(H)/H$) serves the role of the low-field Hall coefficient and should be compared with the
prediction of the models. If this point of view is accepted, then for $H \| ab$ the shape and evolution of the
$d\rho_H(H)/dH$ curves (Fig. \ref{drHHYb}(b)) suggest that possibly the composite fermion model of the QCP is more
relevant to the case of YbAgGe, although for $H \| c$ the shape and evolution of the $d\rho_H(H)/dH$ curves (Fig.
\ref{drHHYbC}(b)) are at variance with the simple theoretical views. The lack of the $T < 0.4$ K data and an
absence of more detailed, realistic-Fermi-surface-tailored, model do not allow us to choose the physical picture
of the field-induced QCP in YbAgGe unambiguously.

\begin{acknowledgments}
Ames Laboratory is operated for the U.S. Department of Energy by Iowa State University under Contract No.
W-7405-Eng.-82. This work was supported by the Director for Energy Research, Office of Basic Energy Sciences. We
thank H. B. Rhee for assistance in growing some of the additional crystals needed for this work.
\end{acknowledgments}

\clearpage

\clearpage

\clearpage

\clearpage

\clearpage

\clearpage

\clearpage

\clearpage

\clearpage

\clearpage

\clearpage

\end{document}